\documentstyle[12pt,epsf]{article}
\newlength{\captionwidth}
\renewcommand{\thefootnote}{\fnsymbol{footnote}}
\newcommand{\lsim}{\;\rlap{\lower 3.5 pt \hbox{$\mathchar \sim$}} \raise 1pt
 \hbox {$<$}\;}
\newcommand{\qsla}{q\hspace{-.5em/\hspace{.5em}}}
\newcommand{\dd}{{\rm d}}
\newcommand{\zpade}{$z$-Pad\'e}
\newcommand{\wpade}{$\omega$-Pad\'e}

\begin{document}    

\title{\vskip-3cm{\baselineskip14pt
\centerline{\normalsize\hfill BUTP-99/10}
\centerline{\normalsize\hfill TTP99--25}
\centerline{\normalsize\hfill hep-ph/9906273}
\centerline{\normalsize\hfill June 1999}
}
\vskip.7cm
Second Order QCD Corrections to $\Gamma(t \to Wb)$
\vskip.5cm
}
\author{{K.G. Chetyrkin}$^{a,}$\thanks{Permanent address:
    Institute for Nuclear Research, Russian Academy of Sciences,
    60th October Anniversary Prospect 7a, Moscow 117312, Russia.}
  \,,
R. Harlander$^a$,
T. Seidensticker$^a$,\\ and 
M. Steinhauser$^b$
  \\[3em]
 {\it (a) Institut f\"ur Theoretische Teilchenphysik,}\\
  {\it Universit\"at Karlsruhe, D-76128 Karlsruhe, Germany}
  \\[.5em]
  {\it (b) Institut f\"ur Theoretische Physik,}\\ 
  {\it Universit\"at Bern, CH-3012 Bern, Switzerland}
}
\date{}
\maketitle

\begin{abstract}
  \noindent Corrections of ${\cal O}(\alpha_s^2)$ to the decay of the
  top quark into a $W$ boson and a bottom quark are calculated. The
  method is based on an expansion of the top quark propagator for small
  external momentum, $q$, as compared to the top quark mass, $M_t$. The
  physical point $q^2 = M_t^2$ is reached through Pad\'e approximations.
  The described method allows to take effects induced by a finite $W$
  boson mass into account.  The numerical relevance of the result is
  discussed.  Important cross-checks against recent results for the
  decay rate $b\to ul\bar\nu$ and the two-loop QED corrections to 
  $\mu$-decay are performed.
\end{abstract}

\thispagestyle{empty}
\newpage
\setcounter{page}{1}


\renewcommand{\thefootnote}{\arabic{footnote}}
\setcounter{footnote}{0}

\section{Introduction}

The top quark is the so far heaviest observed particle of the Standard Model 
of elementary particle physics. Its total width, $\Gamma_t$, is to a good 
approximation proportional to the third power of its mass and is much
larger than $\Lambda_{\rm QCD}$, the typical scale of non-perturbative effects
in QCD. Therefore it is possible to treat the top quark almost as a free
particle and to apply perturbative methods to evaluate the quantum corrections 
to its decay process~\cite{topdec}.

In the minimal Standard Model the dominant decay mode of the top quark is the 
one into a bottom quark and a $W$ boson. It is important to predict the
corresponding decay width accurately in order to be sensitive to exotic
processes, which may occur in supersymmetric models, for example.

The first order QCD corrections have been evaluated in analytical form
some time ago~\cite{JezKue89} and amount to approximately $-10$\%. The
electroweak corrections are small~\cite{DenSac91Eiletal91} and turn out
to be $\approx2$\% for a Higgs mass around $100$~GeV (see,
e.g.,~\cite{toprep98}).

The expected precision for measurements of $\Gamma_t$ by a future $e^+e^-$ 
machine like the Next-Linear-Collider (NLC) is of the same order of magnitude 
as the corrections of ${\cal O}(\alpha_s)$ \cite{toprep98}. This makes it 
desirable to control also the next-to-leading corrections induced by the 
strong interaction.

In fact, the QCD corrections of ${\cal O}(\alpha_s^2)$ have already been 
considered in~\cite{CzaMel98}. This calculation was based on an expansion 
of the vertex diagrams in the quantities $1 - M_b^2/M_t^2$ and 
$1 - 3 M_b^2/M_t^2$, respectively. Although this expansion parameter is not 
small at all, the approach led to reliable results after including many terms 
into the analysis, choosing proper variables, and carefully investigating 
potentially large contributions.

The aim of this paper is, on the one hand, to have an independent check
of the results of~\cite{CzaMel98}, using a rather complementary method.
On the other hand, our approach will allow us to additionally account for
a finite $M_W$ boson mass.

\begin{figure}
\begin{center}
\leavevmode
\epsffile[133 600 485 728]{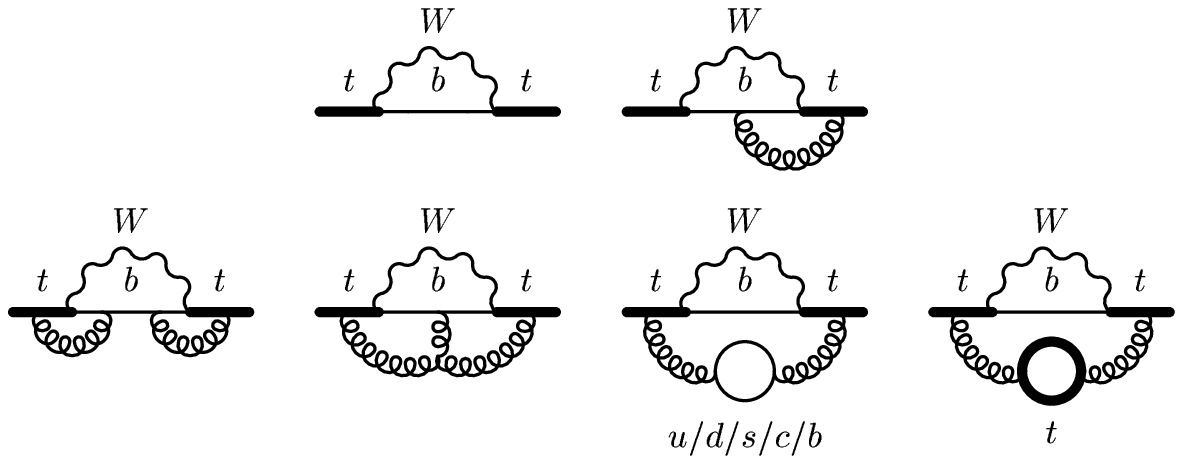}
\end{center}
\parbox{\captionwidth}{
\caption{\label{fig:twb}Sample diagrams for the top quark self energy.}
}
\end{figure}

The method presented in this paper is as follows.  In contrast
to~\cite{CzaMel98} we compute propaga\-tor-type diagrams contributing to
the top quark selfenergy with external momentum $q$ in terms of an
expansion around $q^2/M_t^2=0$. Some sample diagrams are pictured in
Fig.~\ref{fig:twb}.  The imaginary part combined with the wave function
renormalization of the top quark and evaluated at the physical point
$q^2 = M_t^2$ directly leads to the decay rate. It arises from cuts
where the $W$ boson, the bottom quark and, at higher orders, also gluons
and other light quarks are involved.  The limit $q^2\to M_t^2$ is taken
after performing a Pad\'e approximation. The results for $M_W=0$ will be
shown to be in perfect agreement with the ones of \cite{CzaMel98} which
justifies both the method of \cite{CzaMel98} and the one of the present
paper. The subleading terms in $M_W^2$ turn out to be numerically small.

The calculation once again demonstrates the power of expansion
techniques and their computer implementations in multi-loop
calculations. The analyticity properties of the approximated function
guarantee reasonable convergence to the exact result, especially if the
obtained series is further subject to advanced methods like Pad\'e
approximation.

The paper is organized as follows: In Section~\ref{sec:method} the
method is described. In Section~\ref{sec:2loop}, the results obtained at
${\cal O}(\alpha_s)$ are discussed in more detail. The comparison with
the analytical result demonstrates the reliability of our method.
Section~\ref{sec:3loop} deals with the computation of the second order
QCD corrections where also the effects of a finite $W$ boson mass are
taken into account. Important cross-checks with recent results for
$\Gamma(b\to ul\bar\nu)$ and $\Gamma(\mu\to e\nu_\mu\bar\nu_e)$ 
are carried out in Section~\ref{sec:bulnu}.


\section{\label{sec:method}Method of the calculation}

The exact evaluation of the Feynman diagrams contributing to
$\Gamma(t\to Wb)$ at order $\alpha_s^2$ is currently not
available. However, it is promising to apply the methods of asymptotic
expansion (see, e.g., \cite{HarSte98} and references therein) in the
limit
\begin{eqnarray}
M_t^2 \gg q^2 \gg M_W^2 \gg M_b^2 =0\,.
\label{eq:hier}
\end{eqnarray}
For the ${\cal O}(\alpha_s)$ corrections in the limit $M_W=0$,
a similar approach has already been used in~\cite{CzaMel98_2}. There it
was possible to resum the series at $q^2 = M_t^2$ which reproduces the
analytical result. At ${\cal O}(\alpha_s^2)$, however, instead of an
explicit resummation we will perform a Pad\'e approximation in order to reach
the physical point $q^2 = M_t^2$ \cite{FleTar94,BroFleTar93}.

Before going into details, let us introduce the notation. The inverse quark
propagator is denoted by
\begin{eqnarray}
\left( S_F^0(q) \right)^{-1} &=& i\left(m_t^0 \left( 1 - \Sigma_S^0 \right) 
  - \qsla\!\!\! \left( 1 + \Sigma_V^0 \right)\right).
\label{eq:sfinv}
\end{eqnarray}
Both $\Sigma_S^0$ and $\Sigma_V^0$ are functions of the
external momentum $q$ and the bare mass $m^0_t$ of the top quark. In our
case they further depend on the bare strong coupling constant
$\alpha_s^0$ and the $W$ boson mass $M_W$, and are proportional to
the Fermi coupling constant, $G_F$.  $S_F^0$ will become finite
after renormalizing the parameters $m_t^0 = Z_m^{OS} M_t$ and
$\alpha_s^0 = Z_g^2 \alpha_s$, and taking into account the wave function
renormalization:
\begin{eqnarray}              
S_F^{OS} &=& \frac{1}{Z_2^{OS}} S_F^0.
\end{eqnarray}
$Z_m^{OS}$, $Z_2^{OS}$ and $Z_g$ denote the renormalization
constants. $Z_m^{OS}$ and $Z_2^{OS}$
will be taken in the on-shell scheme, whereas $Z_g$ is in the 
$\overline{\rm MS}$ scheme. $Z_2^{OS}$ is defined by the condition 
\begin{eqnarray}
S_F^{OS}(q) &\longrightarrow& \frac{-i}{M_t - \qsla} 
+ \mbox{terms regular for} \,\, q^2 \to M_t^2
\,.
\label{eq:sfos}
\end{eqnarray}

In our approach we are actually dealing with two different masses for
the top quark in intermediate steps: an ``external'' one ($q^2$) and an
``internal'' one ($M_t^2$). Applying the optical theorem, the decay
rate of the top quark will therefore be written as
\begin{eqnarray}
\Gamma(t \rightarrow Wb) &=& 
  ( 2 M_t \, \mbox{Im} [ z\,S_V^{OS} - S_S^{OS} ])\bigg|_{z=1},
\label{eq:gamtwb}
\end{eqnarray}
where
\begin{eqnarray}
S_S^{OS} &=& Z_2^{OS} Z_m^{OS} \left( 1 - \Sigma_S^0 \right)
\,,
\nonumber
\\
\label{eq:sssv}
S_V^{OS} &=& Z_2^{OS}\left(1+\Sigma_V^0\right)
\end{eqnarray}
are functions of the variable
\begin{equation}
z = {q^2\over M_t^2}\,.
\end{equation}
All relevant diagrams will be calculated in terms of expansions around $z = 0$,
and the limit $z \rightarrow 1$ will be applied only in the very end.
Therefore, while $Z_m^{OS}$ can be taken at $z = 1$,
we also need to express $Z_2^{OS}$ in terms of an expansion around $z =
0$. This 
is most conveniently done by translating condition (\ref{eq:sfos}) into 
\begin{equation}
Z_2^{OS} = \left[ 1+\Sigma_V^0 + 
  2\, {\dd\over \dd z}\left(Z_m^{OS}\Sigma_S^0 -
  \Sigma_V^0\right)\right]^{-1}\,. 
\label{eq:Z2os}
\end{equation}
From (\ref{eq:gamtwb}) and (\ref{eq:sssv}) it is clear that for our
purpose it suffices to know only the pure QCD corrections of $Z_2^{OS}$
and $Z_m^{OS}$ up to ${\cal O}(\alpha_s^2)$.  For $z=1$
these quantities were computed up to this order in~\cite{BroGraSch91}
and~\cite{GraBroGraSch90}, respectively.

Note that in a calculation where the quantities are evaluated
on-shell, i.e. at $q^2 = M_t^2$, infra-red singularities appear in
intermediate steps. In contrast, in Eq.~(\ref{eq:sssv}) all
functions on the r.h.s.\ are defined through the expansion for $z\to 0$
and thus are infra-red safe.

At this point a comment on the extraction of the values for $z=1$ is in
order.  Actually Eqs.~(\ref{eq:gamtwb}) and~(\ref{eq:Z2os}) are not
unique as it is possible to derive slightly different equations for 
$Z_2^{OS}$ and $\Gamma(t\to Wb)$, which differ by relative factors of $z$. 
In the limit $z\to1$ all of them are equivalent. The results we
obtained by using two more variants of Eqs.~(\ref{eq:gamtwb})
and~(\ref{eq:Z2os}) are consistent with the ones which will be discussed below.
We decided to use the formulae shown above because the ${\cal O}(\alpha_s)$
corrections are recovered with the highest accuracy. 

In order to obtain reliable results it is necessary to evaluate as many
terms as possible in the expansion parameter $z$.  The exact resummation
of the series in $z$ seems to be excluded. Instead, we apply a Pad\'e
approximation which means that we reexpress the resulting polynomial in
terms of a rational function:
\begin{eqnarray}
[m/n](z) &=& \frac{a_0+a_1 z+\ldots+a_m z^m}
                       {1  +b_1 z+\ldots+b_n z^n}\,.
\label{eq:padedef}
\end{eqnarray}
Its Taylor series is required to coincide with the original polynomial
up to the order $m+n$. For later convenience we define the short hand
notation $[m/n] \equiv [m/n](1)$. The stability of the Pad\'e
approximants upon variation of $m$ and $n$ will indicate the uncertainty
of the approximation (see below).

In addition, it may be promising to apply a conformal mapping
\cite{FleTar94}
\begin{eqnarray}
z &=& \frac{4\omega}{(1+\omega)^2}
\label{eq:confmap} 
\end{eqnarray}
and to perform the Pad\'e approximation in the variable $\omega$.
The complex $z$-plane is mapped into the interior of the unit circle in
the $\omega$-plane, and the relevant point $z = 1$ goes to $\omega = 1$.
This conformal mapping is motivated by the observation that the application
of a Pad\'e approximation relies heavily on analytic properties. The function
on the r.h.s.\ of Eq.~(\ref{eq:gamtwb}) (without the limit $z \to 1$) will
develop a branch cut along the real $z$-axis starting from $z = 1$. This
branch cut is mapped through Eq.~(\ref{eq:confmap}) onto the unit circle. Thus
by applying Eq.~(\ref{eq:confmap}) we enlarge the range of convergence for 
the terms we got in the limit $z \to 0$.    

Since both methods described above appear to be rather natural, any of them 
will be used to derive an estimate on the exact result. For convenience, let us
denote the results obtained through Pad\'e approximation in the variable 
$z$ by ``\zpade s'', the ones where the Pad\'e approximation is performed in
$\omega$ by ``\wpade s''. The central values and the estimated uncertainty will
be extracted from Pad\'e results $[m/n]$ with $m+n$ not too small and 
$|m-n|\le 2$. The central value is obtained by averaging the Pad\'e results
and the uncertainty is given by the maximum deviation from the central value.
The error estimation is therefore rather conservative.

Some Pad\'e approximants develop poles inside the unit circle
($|z|\le1$ and $|\omega|\le1$, respectively). In general we will
discard such results in the following. In some cases, however, the
pole coincides with a zero of the numerator up to several digits
accuracy, and these Pad\'e approximations will be taken into account
for the estimation of the actual results. To be precise: in addition
to the Pad\'e results without any poles inside the unit circle, we
will use the ones where the poles are accompanied by zeros within a
circle of radius 0.01, and the distance between the pole and the
physically relevant point $q^2/M_t^2=1$ is larger than 0.1.

Concerning the dependence on the strong gauge parameter $\xi$ in
Eq.~(\ref{eq:gamtwb}), it only drops out after summing infinitely many
terms in the expansion around $z=0$ and setting $z=1$.
Since we are only dealing with a limited number of terms, our
approximate results will still depend on the choice of $\xi$ even after
taking $z \to 1$.  It is clear that with extreme values of $\xi$ almost
any number could be produced. Thus the question arises which value of
$\xi$ should be assumed in order to arrive at a reliable prediction for
$\Gamma(t\to Wb)$.

At ${\cal O}(\alpha_s)$ the whole calculation can be performed for
arbitrary gauge parameter without any difficulties. This allows for a
detailed study of the residual $\xi$ dependence. At ${\cal
O}(\alpha_s^2)$ only the first few terms could be evaluated for general
$\xi$ which does not allow for extensive studies. In order to arrive
at a reasonable number of terms in the expansion around $z=0$ it is
necessary to set $\xi$ to some definite value from the very beginning.
The behaviour of the diagrams at ${\cal O}(\alpha_s)$ will serve as an
indication for the optimal choice of this value in the analysis at 
${\cal O}(\alpha_s^2)$.

Concerning the electroweak gauge parameter $\xi_W$, all results that will be
quoted in the following have been obtained in unitary gauge, where the
$W$ propagator is given by
\begin{equation}
  D_W(p) = {-i\over M_W^2 - p^2}\left(-g_{\mu\nu} + {p_\mu p_\nu\over
    M_W^2}\right)\,.
\end{equation}
Nevertheless, the leading terms in $M_W$ have also been computed in an
arbitrary covariant gauge. They are obtained by
replacing the $W$ boson by a Goldstone boson with the propagator simply
given by
\begin{equation}
D_\Phi(p) = {-i\over \xi_W M_W^2 - p^2} \stackrel{M_W \to
  0}{\longrightarrow} {i\over p^2}\,.
\end{equation}
The independence of $\xi_W$ is then manifest already at this point.


\section{\label{sec:2loop}First order QCD corrections}

In this section we will investigate the ${\cal O}(\alpha_s)$ corrections 
and compare the exact result~\cite{JezKue89} to the approximation obtained
by the method described above.

It is convenient to decompose the decay rate of the top quark
into a $W$ boson and a bottom quark in the following way:
\begin{eqnarray}\label{eq:coldec}
\Gamma(t\to bW) &=& \Gamma_0\left[
  A^{(0)}
  +\frac{\alpha_s}{\pi} C_F A^{(1)}
  +\left(\frac{\alpha_s}{\pi}\right)^2 A^{(2)}
  +\ldots
\right]
\,,
\end{eqnarray}
where $\Gamma_0=G_FM_t^3|V_{tb}|^2/(8\pi\sqrt{2})$,
$A^{(0)}=1-3M_W^4/M_t^4+2M_W^6/M_t^6$, $C_F = 4/3$, and
$V_{tb}$ is the CKM matrix element for $t\to b$ transitions.
The running coupling $\alpha_s$ is defined with six active flavours.

The one-loop correction is known in analytical form since quite some 
time~\cite{JezKue89}. Expanded in terms of $M_W/M_t$ it reads:
\begin{eqnarray}
A^{(1)} &=&
\frac{5}{4}
-\frac{\pi^2}{3}
+\frac{3}{2}\frac{M_W^2}{M_t^2}
+\frac{M_W^4}{M_t^4}
  \left( -6 +\pi^2 
    -\frac{3}{2}\ln\frac{M_t^2}{M_W^2}\right)
\nonumber\\&&\mbox{}
+\frac{M_W^6}{M_t^6}
  \left( \frac{46}{9} -\frac{2}{3}\pi^2 
    +\frac{2}{3}\ln\frac{M_t^2}{M_W^2}\right)
+{\cal O}\left(\frac{M_W^8}{M_t^8}\right)
\,.
\label{eq:a1}
\end{eqnarray}
The approximation $M_W=0$ induces an error of roughly 22\%.
This reduces to approximately 4\% if the quadratic mass corrections
are included and is completely negligible if all the
terms displayed in Eq.~(\ref{eq:a1}) are taken into account.

For clarity, let us apply our method to these lowest order terms and see
how the results compare to $A^{(0)}$ and $A^{(1)}$ above. While $A^{(0)}$ 
is reproduced exactly, the imaginary part of the small-momentum expansion 
for the two-loop (${\cal O}(\alpha_s)$) diagrams (an example is shown in 
Fig.~\ref{fig:twb}) reads:
\begin{eqnarray}
\lefteqn{A^{(1)}_{exp} =}\nonumber\\&&\mbox{} 
  - {19 \over 12} - {29 \over 144} \, z - {23 \over 240} \, z^2 
  - {61 \over 1200} \, z^3 - {151 \over 5040} \, z^4 
  - {449 \over 23520} \, z^5 - {13 \over 1008} \, z^6
\nonumber\\&&\mbox{} 
  - {827 \over 90720} \, z^7 - {529 \over 79200} \, z^8 
\nonumber\\&&\mbox{}
  + \xi \, \bigg( {5 \over 12} - {7 \over 48} \, z - {1 \over 16} \, z^2
     - {3 \over 80} \, z^3 - {43 \over 1680} \, z^4 - {3 \over 160} \, z^5 
     - {29 \over 2016} \, z^6
\nonumber\\&&\mbox{\hspace{2em}} 
     - {23 \over 2016} \, z^7 
     - {49 \over 5280} \, z^8 \bigg)
\nonumber\\&&\mbox{}
  + \frac{M_W^2}{M_t^2z} \, \bigg[ {3 \over 2} + {1 \over 18} \, z 
     - {1 \over 72} \, z^2 - {7 \over 600} \, z^3 - {7 \over 900} \, z^4    
     - {23 \over 4410} \, z^5 - {17 \over 4704} \, z^6 
\nonumber\\&&\mbox{\hspace{2em}}
     - {47 \over 18144} \, z^7 - {31 \over 16200} \, z^8 \bigg]
\nonumber\\&&\mbox{}
  + \frac{M_W^4}{M_t^4z^2} \, \bigg[ {1 \over 4} 
     + {13 \over 3} \, z - {383 \over 720} \, z^2 - {79 \over 720} \, z^3
     - {313 \over 8400} \, z^4 - {403 \over 25200} \, z^5 
     - {557 \over 70560} \, z^6 
\nonumber\\&&\mbox{\hspace{2em}}
     - {151 \over 35280} \, z^7 - {2477 \over 997920} \, z^8
\nonumber\\&&\mbox{\hspace{2em}}
     + \xi \, \bigg( {5 \over 4} - {5 \over 2} \, z + {37 \over 80} \, z^2
         + {3 \over 16} \, z^3 + {61 \over 560} \, z^4 + {41 \over 560} \, z^5
         + {179 \over 3360} \, z^6
\nonumber\\&&\mbox{\hspace{4em}} 
         + {137 \over 3360} \, z^7 
         + {239 \over 7392} \, z^8 \bigg)
\nonumber\\&&\mbox{\hspace{2em}}
     + l_{tW} \, \bigg( {3 \over 4} - {9 \over 4} \, z
     - {3 \over 4} \, \xi + {3 \over 4} \, z \, \xi \bigg) \bigg]
     + {\cal O} \left( \frac{M_W^6}{M_t^6} \right)\,,
\label{eq:aexp}
\end{eqnarray}
with $l_{tW}=\ln M_t^2/M_W^2$. (The coefficient of $z^n$ will be called the 
``$n^{th}$ moment'' in the following.) Note that the factors $1/z$ and
$1/z^2$ in front of the quadratic and quartic terms in $M_W$ are
irrelevant for the subsequent Pad\'e procedure.

The procedure described above is applied to each coefficient of
$M_W^2/M_t^2$ separately. As already noted, $A^{(1)}_{exp}$ still
depends on the QCD gauge parameter, $\xi$, appearing in the gluon
propagator $i(-g^{\mu\nu}+\xi q^\mu q^\nu/q^2)/(q^2+i\epsilon)$.  Thus
also the Pad\'e approximations $[m/n]$ will show a dependence on
$\xi$.  It is clear that for large absolute values of $\xi$ the
quantities $[m/n]$ get dominated by them and any predictive power is
lost.  In Table~\ref{tab:oasxi} several \zpade s are evaluated for the
leading order coefficient ($M_W=0$). The gauge parameter is varied
from $\xi=-2$ to $\xi=+2$.  Pad\'e results which develop poles for
$|z|\le1$ are in general represented by a dash. However, if an
approximate cancellation with a zero from the numerator takes place
(see the discussion above), they are marked by a star ($\star$).  The
exact result is reproduced with a fairly high accuracy for almost all
values of $\xi$ under consideration.  Nevertheless, the value for
$\xi=0$ is closest to the exact result, and the variation of the
Pad\'e approximants appears to be very small for this particular
choice of $\xi$.  Based on this observation, we decide to perform the
three-loop analysis by setting $\xi=0$. This has the additional
advantage that the evaluation of the Feynman diagrams is much simpler
than for non-zero values of $\xi$.

{\footnotesize
\begin{table}[ht]
\begin{center}
\begin{tabular}{|l|l||r|r|r|r|r|r|r|}
\hline
input
 & 
P.A.
 & 
 $ \xi = -2 $ 
 & 
 $ \xi = -1 $ 
 & 
 $ \xi = -1/2 $ 
 & 
 $ \xi = 0 $ 
 & 
 $ \xi = 1/2 $ 
 & 
 $ \xi = 1 $ 
 & 
 $ \xi = 2 $ 
\\\hline
6
 & 
[3/2]
 & 
 $ -2.111 $ 
 & 
 --- 
 & 
 $ -2.058 $ 
 & 
 $ -2.023 $ 
 & 
 $ -1.990 $ 
 & 
 $ -1.957 $ 
 & 
 $ -1.893 $ 
\\
6
 & 
[2/3]
 & 
 $ -2.112 $ 
 & 
 $ -2.052 $ 
 & 
 $ -2.058 $ 
 & 
 $ -2.023 $ 
 & 
 $ -1.990 $ 
 & 
 $ -1.963 $ 
 & 
 $ -1.887 $ 
\\\hline
7
 & 
[4/2]
 & 
 $ -2.121 $ 
 & 
 --- 
 & 
 $ -2.058 $ 
 & 
 $ -2.027 $ 
 & 
 $ -1.999 $ 
 & 
 $ -1.972 $ 
 & 
 $ -1.919 $ 
\\
7
 & 
[3/3]
 & 
 $ -2.120 $ 
 & 
 --- 
 & 
 $ -2.058 $ 
 & 
 $ ^{(\star)} -2.025 $ 
 & 
 $ -2.008 $ 
 & 
 $ -1.980 $ 
 & 
 $ -1.928 $ 
\\
7
 & 
[2/4]
 & 
 $ -2.126 $ 
 & 
 --- 
 & 
 $ -2.058 $ 
 & 
 $ -2.027 $ 
 & 
 $ -2.000 $ 
 & 
 $ -1.979 $ 
 & 
 $ -1.912 $ 
\\\hline
8
 & 
[4/3]
 & 
 $ -2.117 $ 
 & 
 --- 
 & 
 --- 
 & 
 $ -2.033 $ 
 & 
 $ -2.011 $ 
 & 
 $ -1.990 $ 
 & 
 $ -1.949 $ 
\\
8
 & 
[3/4]
 & 
 $ -2.117 $ 
 & 
 $ -1.694 $ 
 & 
 $ -2.058 $ 
 & 
 $ -2.033 $ 
 & 
 $ -2.011 $ 
 & 
 $ -1.993 $ 
 & 
 $ -1.948 $ 
\\\hline
9
 & 
[5/3]
 & 
 $ -2.112 $ 
 & 
 $ -2.063 $ 
 & 
 $ ^{(\star)} -2.059 $ 
 & 
 $ -2.034 $ 
 & 
 $ -2.016 $ 
 & 
 $ -1.998 $ 
 & 
 $ -1.963 $ 
\\
9
 & 
[4/4]
 & 
 $ -2.105 $ 
 & 
 $ -2.061 $ 
 & 
 $ ^{(\star)} -2.059 $ 
 & 
 $ -2.034 $ 
 & 
 $ -2.040 $ 
 & 
 $ -2.001 $ 
 & 
 $ -1.967 $ 
\\
9
 & 
[3/5]
 & 
 $ -2.114 $ 
 & 
 $ -2.064 $ 
 & 
 $ ^{(\star)} -2.059 $ 
 & 
 $ -2.034 $ 
 & 
 $ -2.016 $ 
 & 
 $ -2.001 $ 
 & 
 $ -1.961 $ 
\\\hline
\multicolumn{2}{|c||}{exact:}
 & 
\multicolumn{7}{|c|}{ $ -2.040 $ }
\\
\hline
\end{tabular}

\parbox{\captionwidth}{
\caption{\label{tab:oasxi}
  \zpade s for $A^{(1)}\big|_{M_W=0}$ for different values of $\xi$. The
  first column indicates the number of terms in $z$ from
  (\ref{eq:aexp}) that were used as input.  }}
\end{center}
\end{table}
}

Let us now discuss the sub-leading terms in $M_W^2$. Table~\ref{tab:oasmw}
lists several Pad\'e approximations for the coefficients of
$(M_W^2/M_t^2)^n$ ($n=0,1,2$) in the case of $A^{(1)}$. $\xi$ has been
set to zero, as it is suggested by the discussion above. For each
coefficient, the Pad\'e approximations have been performed in the
variable $z$ as well as in the variable $\omega$. The $z$- and \wpade s 
are indicated by a $z$ and an $\omega$, respectively, in the second line of
Table~\ref{tab:oasmw}. This notation will be used throughout the paper. 
The \zpade s for the values
of the $(M_W^2/M_t^2)^0$-term coincide with those for $\xi=0$ of
Table~\ref{tab:oasxi}, of course.  Concerning the power-suppressed
terms, again the higher order Pad\'e approximants agree with the exact
results to an impressive accuracy.
The logarithm of Eq.~(\ref{eq:a1}) is reproduced exactly after taking
into account the first two terms in the expansion (\ref{eq:aexp}) and
setting $z = 1$.
{\footnotesize
\begin{table}[ht]
\begin{center}
\begin{tabular}{|l|l||r|r||r|r||r|r|r|}
\hline
\multicolumn{2}{|c||}{ }
 & 
\multicolumn{2}{|c||}{ $ M_W^0 $ }
 & 
\multicolumn{2}{|c||}{ $ M_W^2 $ }
 & 
\multicolumn{2}{|c|}{ $ M_W^4 $ }
\\\hline
input
 & 
P.A.
 & 
\multicolumn{1}{|c|}{ $ z $ }
 & 
\multicolumn{1}{|c||}{ $ \omega $ }
 & 
\multicolumn{1}{|c|}{ $ z $ }
 & 
\multicolumn{1}{|c||}{ $ \omega $ }
 & 
\multicolumn{1}{|c|}{ $ z $ }
 & 
\multicolumn{1}{|c|}{ $ \omega $ }
\\\hline
6
 & 
[3/2]
 & 
 $ -2.023 $ 
 & 
 $ -2.022 $ 
 & 
 $ 1.506 $ 
 & 
 $ 1.508 $ 
 & 
 $ 3.875 $ 
 & 
 --- 
\\
6
 & 
[2/3]
 & 
 $ -2.023 $ 
 & 
 $ -2.051 $ 
 & 
 --- 
 & 
 $ 1.548 $ 
 & 
 $ 3.878 $ 
 & 
 --- 
\\\hline
7
 & 
[4/2]
 & 
 $ -2.027 $ 
 & 
 $ ^{(\star)} -2.009 $ 
 & 
 $ ^{(\star)} 1.507 $ 
 & 
 $ ^{(\star)} 1.518 $ 
 & 
 $ 3.871 $ 
 & 
 --- 
\\
7
 & 
[3/3]
 & 
 $ ^{(\star)} -2.025 $ 
 & 
 $ -2.035 $ 
 & 
 --- 
 & 
 $ 1.501 $ 
 & 
 $ 3.872 $ 
 & 
 --- 
\\
7
 & 
[2/4]
 & 
 $ -2.027 $ 
 & 
 $ -2.040 $ 
 & 
 --- 
 & 
 $ 1.507 $ 
 & 
 $ 3.874 $ 
 & 
 --- 
\\\hline
8
 & 
[4/3]
 & 
 $ -2.033 $ 
 & 
 $ -2.035 $ 
 & 
 $ 1.502 $ 
 & 
 $ 1.502 $ 
 & 
 $ 3.870 $ 
 & 
 --- 
\\
8
 & 
[3/4]
 & 
 $ -2.033 $ 
 & 
 $ -2.035 $ 
 & 
 --- 
 & 
 $ 1.502 $ 
 & 
 $ 3.870 $ 
 & 
 --- 
\\\hline
9
 & 
[5/3]
 & 
 $ -2.034 $ 
 & 
 $ -2.039 $ 
 & 
 $ 1.502 $ 
 & 
 $ 1.500 $ 
 & 
 $ 3.870 $ 
 & 
 --- 
\\
9
 & 
[4/4]
 & 
 $ -2.034 $ 
 & 
 $ ^{(\star)} -2.035 $ 
 & 
 $ 1.502 $ 
 & 
 $ ^{(\star)} 1.501 $ 
 & 
 $ 3.870 $ 
 & 
 --- 
\\
9
 & 
[3/5]
 & 
 $ -2.034 $ 
 & 
 $ -2.037 $ 
 & 
 $ 1.502 $ 
 & 
 $ 1.499 $ 
 & 
 $ 3.870 $ 
 & 
 --- 
\\\hline
\multicolumn{2}{|c||}{exact:}
 & 
\multicolumn{2}{|c||}{ $ -2.040 $ }
 & 
\multicolumn{2}{|c||}{ $ 1.500 $ }
 & 
\multicolumn{2}{|c|}{ $ 3.870 $ }
\\
\hline
\end{tabular}

\parbox{\captionwidth}{
\caption[]{\label{tab:oasmw}
  Pad\'e approximations for the power-suppressed terms of $A^{(1)}$,
  computed for $\xi=0$.}}
\end{center}
\end{table}
}

Also for the power-suppressed terms an analysis concerning the $\xi$
dependence has been performed. While the quadratic terms in $M_W$ do not
depend on $\xi$ at all (see Eq.~(\ref{eq:aexp})), the
results for the terms of order $M_W^4/M_t^4$ become significantly worse
once $\xi$ is different from zero.  This also supports the choice of
Feynman gauge at three loops.

Taking only those results of Table~\ref{tab:oasmw} into account where
eight or more input terms enter we may finally deduce our approximation for 
the ${\cal O}(\alpha_s)$ corrections to the decay rate (we adopt the
notation $-2.035(4) \equiv -2.035\pm 0.004$, etc.):
\begin{eqnarray}\label{eq:res1a}
A^{(1)} &=& -2.035(4) + 1.501(2)\,\frac{M_W^2}{M_t^2} +
\frac{M_W^4}{M_t^4}\left(3.8700(5) - {3\over 2}\,\ln{M_t^2\over M_W^2}\right)
+\ldots\,.
\end{eqnarray}
The agreement of Eq.~(\ref{eq:res1a}) with the exact results quoted in the
last line of Table~\ref{tab:oasmw} obviously supports the underlying method.

In the three-loop case we could obtain the small-momentum expansion
up to $z^7$ for $M_W=0$ and up to $z^6$ for the coefficients of
$M_W^2/M_t^2$ and the $M_W^4/M_t^4$. Therefore, the final number at
three-loop level will be based on Pad\'es built out of seven and eight
moments for the leading term, and six and seven moments for the
sub-leading terms in $M_W$. At ${\cal O}(\alpha_s)$ this reduced number
of input terms changes the result from the one in (\ref{eq:res1a}) to
\begin{eqnarray}
A^{(1)} &=& - 2.03(2) + 1.51(4)\,{M_W^2\over M_t^2} + 
{M_W^4\over M_t^4}\left(3.874(4) - {3\over 2}\ln {M_t^2\over M_W^2}\right)
+\ldots\,.
\label{eq:res1b}
\end{eqnarray}
which is a bit worse than (\ref{eq:res1a}), but still sufficiently accurate.
This suggests that the number of available moments at order $\alpha_s^2$ 
should be sufficient to arrive at a reasonable estimate.


\section{\label{sec:3loop}Second order QCD corrections}

Let us use the experience gained in the previous section to 
obtain predictions for $\Gamma(t\to Wb)$ at order $\alpha_s^2$.

\subsection{General remarks}

It is convenient to
decompose the decay rate according to the colour structure:
\begin{eqnarray}
A^{(2)} &=&
C_F^2 A_A^{(2)}
+C_AC_F A_{NA}^{(2)}
+C_F T n_l A_{l}^{(2)}
+C_F T A_{F}^{(2)}
\label{eq:2loopdef}
\,,
\end{eqnarray}
where in QCD the colour factors are given by $C_F = 4/3$, $C_A = 3$, and
$T=1/2$. $n_l$ is the number of massless quark flavours and will be set
to $n_l=5$ in the end. $A_A^{(2)}$ corresponds to the abelian part
already present in QED, $A_{NA}^{(2)}$ represents the non-abelian
contribution, and $A_l^{(2)}$ and $A_F^{(2)}$ denote the corrections
involving a second fermion loop with massless and massive quarks,
respectively.  In Fig.~\ref{fig:twb} a representative diagram for each
function is pictured.  The expansion in terms of $M_W^2/M_t^2$ of the
individual contributions to $A^{(2)}$ will be written as
\begin{equation} 
A^{(2)}_i = A^{(2)}_i|_{M_W=0} + {M_W^2\over M_t^2}\, A^{(2)}_i|_{M_W^2} +
{M_W^4\over M_t^4}\,A^{(2)}_i|_{M_W^4} + \ldots\,,
\end{equation}
with $i=A,\,NA,\,l$ and $F$. Note that $A^{(2)}_l|_{M_W=0}$ is known
analytically~\cite{Cza95} and serves as a welcome check for our method.

Whereas at ${\cal O}(\alpha_s)$ the 't~Hooft mass $\mu^2$ drops out (see
Eq.~(\ref{eq:aexp})), it does appear at ${\cal O}(\alpha_s^2)$. We adopted 
the convention $\mu^2 = M_t^2$ throughout the paper. 

There are 60 three-loop diagrams that contribute to $\Gamma(t\to Wb)$.
The practical computation is done with the help of the package {\tt
  GEFICOM}~\cite{geficom}. It uses {\tt QGRAF}~\cite{Nog93} for the
generation of the diagrams and {\tt EXP}~\cite{Sei:dipl} for the
application of the hard mass procedure.  For more details we refer to a
recent review concerned with the automatic computation of Feynman
diagrams~\cite{HarSte98}.  The application of the methods of asymptotic
expansion according to 
Eq.~(\ref{eq:hier}) reduces the practical computation
either to massless three-loop propagator-type diagrams or to products of one-
and two-loop integrals. In the latter case either vacuum graphs or again
massless two-point functions appear. The integrals have been performed with
the help of the packages {\tt MINCER}~\cite{MINCER} and {\tt
MATAD}~\cite{Ste96} based on the symbolic manipulation language 
{\tt FORM}~\cite{FORM}. The results for $\Sigma_S^0$ and $\Sigma_V^0$ 
defined in Eqs.~(\ref{eq:gamtwb}) and~(\ref{eq:Z2os}) at three loops are quite 
lengthy and therefore not listed here. They can be obtained from the authors
upon request.

\subsection{The limit $M_W=0$}

This section is concerned with the second order QCD corrections where
the mass of the $W$ boson is neglected. In this limit a comparison
with~\cite{CzaMel98} can be performed.  The $M_W$-suppressed corrections
will be discussed in the subsequent section.  As already noted,
$A_{A,exp}^{(2)}(z)$ and $A_{NA,exp}^{(2)}(z)$ were computed up to $z^7$
in the case of a massless $W$ boson in Feynman gauge which corresponds
to eight input terms for the Pad\'e approximations.

The fermionic pieces, $A_{l,exp}^{(2)}$ and $A_{F,exp}^{(2)}$, do
not depend on the QCD gauge parameter and are of simpler structure than
$A_{A,exp}^{(2)}$ and $A_{NA,exp}^{(2)}$. For the light-fermion
contribution we thus could evaluate nine terms in the expansion around
$z=0$ for the leading term in $M_W$.

In Table~\ref{tab:oas2allmb0} the results are displayed.
The \zpade s for $A_{F}^{(2)}$ converge very quickly whereas most of the
\wpade s develop poles for $|\omega|\le1$.
As a result we infer
\begin{eqnarray}
A_{F}^{(2)}\bigg|_{M_W=0} &=& -0.06360(1)
\,,
\label{eq:resalF}
\end{eqnarray}
which coincides with the one quoted in~\cite{CzaMel98}. Note, however,
that the magnitude of $A_{F}^{(2)}$ is rather small.

{\footnotesize
\begin{table}[ht]
\begin{center}
\begin{tabular}{|l|l||r|r||r|r||r|r||r|r|}
\hline
\multicolumn{2}{|c||}{ }
 & 
\multicolumn{2}{|c||}{ $ A_{l}^{(2)} $ }
 & 
\multicolumn{2}{|c||}{ $ A_{F}^{(2)} $ }
 & 
\multicolumn{2}{|c||}{ $ A_{A}^{(2)} $ }
 & 
\multicolumn{2}{|c|}{ $ A_{NA}^{(2)} $ }
\\\hline
input
 & 
P.A.
 & 
\multicolumn{1}{|c|}{ $ z $ }
 & 
\multicolumn{1}{|c||}{ $ \omega $ }
 & 
\multicolumn{1}{|c|}{ $ z $ }
 & 
\multicolumn{1}{|c||}{ $ \omega $ }
 & 
\multicolumn{1}{|c|}{ $ z $ }
 & 
\multicolumn{1}{|c||}{ $ \omega $ }
 & 
\multicolumn{1}{|c|}{ $ z $ }
 & 
\multicolumn{1}{|c|}{ $ \omega $ }
\\\hline
6
 & 
[3/2]
 & 
 $ 2.600 $ 
 & 
 $ 2.889 $ 
 & 
 $ -0.06359 $ 
 & 
 --- 
 & 
 $ 2.971 $ 
 & 
 --- 
 & 
 $ -7.637 $ 
 & 
 $ -8.238 $ 
\\
6
 & 
[2/3]
 & 
 $ 2.600 $ 
 & 
 $ 2.994 $ 
 & 
 $ -0.06359 $ 
 & 
 $ -0.06464 $ 
 & 
 $ 2.972 $ 
 & 
 --- 
 & 
 $ -7.639 $ 
 & 
 $ -8.400 $ 
\\\hline
7
 & 
[4/2]
 & 
 $ 2.633 $ 
 & 
 $ 2.920 $ 
 & 
 $ -0.06360 $ 
 & 
 --- 
 & 
 $ 3.041 $ 
 & 
 $ 3.818 $ 
 & 
 $ -7.710 $ 
 & 
 $ -8.212 $ 
\\
7
 & 
[3/3]
 & 
 $ ^{(\star)} 2.592 $ 
 & 
 $ 2.926 $ 
 & 
 $ -0.06360 $ 
 & 
 --- 
 & 
 $ 3.140 $ 
 & 
 --- 
 & 
 $ -7.781 $ 
 & 
 $ -8.218 $ 
\\
7
 & 
[2/4]
 & 
 $ 2.636 $ 
 & 
 $ 2.956 $ 
 & 
 $ -0.06359 $ 
 & 
 --- 
 & 
 $ 3.046 $ 
 & 
 --- 
 & 
 $ -7.719 $ 
 & 
 $ -8.306 $ 
\\\hline
8
 & 
[4/3]
 & 
 $ 2.695 $ 
 & 
 $ ^{(\star)} 2.907 $ 
 & 
 $ -0.06360$ 
 & 
 --- 
 & 
 $ 3.146 $ 
 & 
 --- 
 & 
 $ -7.820 $ 
 & 
 $ ^{(\star)} -8.232 $ 
\\
8
 & 
[3/4]
 & 
 $ 2.696 $ 
 & 
 $ 2.832 $ 
 & 
 $ -0.06360$ 
 & 
 --- 
 & 
 $ 3.146 $ 
 & 
 --- 
 & 
 $ -7.820 $ 
 & 
 $ -8.254 $ 
\\\hline
9
 & 
[5/3]
 & 
 $ 2.708 $ 
 & 
 $ 2.881 $ 
\\
9
 & 
[4/4]
 & 
 $ 2.707 $ 
 & 
 $ 2.892 $ 
\\
9
 & 
[3/5]
 & 
 $ 2.708 $ 
 & 
 $ 2.902 $ 
\\\cline{1-4}
\multicolumn{2}{|c||}{exact:}
 & 
\multicolumn{2}{|c||}{2.859}
\\
\cline{1-4}
\end{tabular}

\parbox{\captionwidth}{
\caption[]{\label{tab:oas2allmb0}
  Pad\'e results at ${\cal O}(\alpha_s^2)$ for $M_W=0$.}}
\end{center}
\end{table}
}

$A_{l}^{(2)}$ behaves similar to the ${\cal O}(\alpha_s)$ corrections.
As expected, the more terms of the expansion in $z$ are included, the better
agreement with the exact result is observed. Furthermore, both \zpade s and 
\wpade s lead to compatible numerical values from which, including the seventh
and eighth moment (``input 8'' and ``input 9'' in
Table~\ref{tab:oas2allmb0}), the following result is deduced:
\begin{eqnarray}
A_{l}^{(2)}\bigg|_{M_W=0} &=& 2.8(1)
\,.
\label{eq:resal1}
\end{eqnarray}
The error is around 4\% and thus roughly as large as the one
in~\cite{CzaMel98}, where the result reads $2.85(7)$. 
Using only the sixth and seventh moment one ends up with
\begin{eqnarray}
A_{l}^{(2)}\bigg|_{M_W=0} &=& 2.8(2)
\,,
\label{eq:resal2}
\end{eqnarray}
where the error is 7\%.
The result of Eqs.~(\ref{eq:resal1}) and~(\ref{eq:resal2}) can also be
compared with the exact number~\cite{Cza95} which reads $2.859\ldots$.

Let us now turn to the abelian and non-abelian parts.  Like in the
one-loop case, the expansion in $z$ is gauge dependent for these
contributions.  Motivated by the observations of
Section~\ref{sec:2loop}, the analysis will be performed by setting
$\xi=0$ from the very beginning.  For non-zero values of $\xi$ some
lower-order Pad\'e approximants will be presented at the end of this
section.

As compared to the fermionic contributions, the spread among the \zpade
s and \wpade s is significantly larger. Moreover, the numbers for the
two approaches are less compatible with each other.  Nevertheless,
following the previously introduced strategy for the extraction of the
central value and the error, we obtain
\begin{eqnarray}
A_{A}^{(2)}\bigg|_{M_W=0} &=& 3.2(6)
\,,
\label{eq:Amw0}
\\
A_{NA}^{(2)}\bigg|_{M_W=0} &=& -8.0(3)
\,.
\label{eq:NAmw0}
\end{eqnarray}

The (fairly conservative) errors are larger than the ones of the results
in~\cite{CzaMel98}, which read $3.5(2)$ and $-8.10(17)$, respectively. The numbers, however, are consistent.

At this point we have confirmed the results of~\cite{CzaMel98} with a
completely independent method. Our calculation can therefore serve as an
important cross check.

Due to the complexity of the intermediate expressions it was impossible
to evaluate eight terms in the expansion for small $z$ using a general
gauge parameter.  In this case, we managed to compute the expansions for
$A_A^{(2)}$ and $A_{NA}^{(2)}$ only up to terms of ${\cal O}(z^3)$.  In
Tables~\ref{tab:Axin} and \ref{tab:NAxin} the gauge parameter is varied
between $\xi=-2$ and $\xi=+2$. For $A_A^{(2)}$ the error bars of
Eq.~(\ref{eq:Amw0}) are conservative enough to cover even most of the
values of Table~\ref{tab:Axin}. The stability of $A_{NA}^{(2)}$ against
variations of $\xi$ is also satisfactory, although many of the values of
Table~\ref{tab:NAxin} are not compatible with (\ref{eq:NAmw0}). However,
this could be traced to the low number of input terms in
Table~\ref{tab:NAxin}.

{\footnotesize
\begin{table}[ht]
\begin{center}
\begin{tabular}{|l|l||r|r|r|r|r|r|r|}
\hline
input
 & 
P.A.
 & 
 $ \xi = -2 $ 
 & 
 $ \xi = -1 $ 
 & 
 $ \xi = -1/2 $ 
 & 
 $ \xi = 0 $ 
 & 
 $ \xi = 1/2 $ 
 & 
 $ \xi = 1 $ 
 & 
 $ \xi = 2 $ 
\\\hline
3
 & 
[2/0]
 & 
 $ 2.790 $ 
 & 
 $ 2.425 $ 
 & 
 $ 2.305 $ 
 & 
 $ 2.226 $ 
 & 
 $ 2.189 $ 
 & 
 $ 2.194 $ 
 & 
 $ 2.329 $ 
\\
3
 & 
[1/1]
 & 
 $ 2.780 $ 
 & 
 --- 
 & 
 $ 2.685 $ 
 & 
 $ 2.508 $ 
 & 
 $ 2.518 $ 
 & 
 $ 2.632 $ 
 & 
 $ 3.254 $ 
\\
3
 & 
[0/2]
 & 
 $ 2.787 $ 
 & 
 $ 2.432 $ 
 & 
 $ 2.329 $ 
 & 
 $ 2.280 $ 
 & 
 $ 2.293 $ 
 & 
 $ 2.375 $ 
 & 
 $ 2.789 $ 
\\\hline
4
 & 
[3/0]
 & 
 $ 2.849 $ 
 & 
 $ 2.501 $ 
 & 
 $ 2.397 $ 
 & 
 $ 2.341 $ 
 & 
 $ 2.332 $ 
 & 
 $ 2.369 $ 
 & 
 $ 2.586 $ 
\\
4
 & 
[2/1]
 & 
 $ ^{(\star)} 2.624 $ 
 & 
 $ 3.268 $ 
 & 
 $ 2.810 $ 
 & 
 $ 2.713 $ 
 & 
 $ 2.725 $ 
 & 
 $ 2.816 $ 
 & 
 $ 3.201 $ 
\\
4
 & 
[1/2]
 & 
 $ ^{(\star)} 2.705 $ 
 & 
 $ 3.320 $ 
 & 
 $ 2.810 $ 
 & 
 $ 2.721 $ 
 & 
 $ 2.739 $ 
 & 
 $ 2.827 $ 
 & 
 $ 3.202 $ 
\\
4
 & 
[0/3]
 & 
 $ 2.847 $ 
 & 
 $ 2.517 $ 
 & 
 $ 2.433 $ 
 & 
 $ 2.411 $ 
 & 
 $ 2.455 $ 
 & 
 $ 2.572 $ 
 & 
 $ 3.047 $ 
\\
\hline
\end{tabular}

\parbox{\captionwidth}{
\caption[]{\label{tab:Axin}
  $\xi$ dependence of $A_{A}^{(2)}|_{M_W=0}$.
  }}
\end{center}
\end{table}
}
{\footnotesize
\begin{table}[ht]
\begin{center}
\begin{tabular}{|l|l||r|r|r|r|r|r|r|}
\hline
input
 & 
P.A.
 & 
 $ \xi = -2 $ 
 & 
 $ \xi = -1 $ 
 & 
 $ \xi = -1/2 $ 
 & 
 $ \xi = 0 $ 
 & 
 $ \xi = 1/2 $ 
 & 
 $ \xi = 1 $ 
 & 
 $ \xi = 2 $ 
\\\hline
3
 & 
[2/0]
 & 
 $ -7.894 $ 
 & 
 $ -6.953 $ 
 & 
 $ -6.571 $ 
 & 
 $ -6.248 $ 
 & 
 $ -5.984 $ 
 & 
 $ -5.779 $ 
 & 
 $ -5.547 $ 
\\
3
 & 
[1/1]
 & 
 $ -8.202 $ 
 & 
 $ -7.455 $ 
 & 
 $ -7.176 $ 
 & 
 $ -6.954 $ 
 & 
 $ -6.783 $ 
 & 
 $ -6.655 $ 
 & 
 $ -6.515 $ 
\\
3
 & 
[0/2]
 & 
 $ -8.078 $ 
 & 
 $ -7.243 $ 
 & 
 $ -6.924 $ 
 & 
 $ -6.669 $ 
 & 
 $ -6.476 $ 
 & 
 $ -6.340 $ 
 & 
 $ -6.219 $ 
\\\hline
4
 & 
[3/0]
 & 
 $ -8.117 $ 
 & 
 $ -7.262 $ 
 & 
 $ -6.915 $ 
 & 
 $ -6.623 $ 
 & 
 $ -6.385 $ 
 & 
 $ -6.202 $ 
 & 
 $ -5.997 $ 
\\
4
 & 
[2/1]
 & 
 $ -8.335 $ 
 & 
 $ -7.670 $ 
 & 
 $ -7.417 $ 
 & 
 $ -7.212 $ 
 & 
 $ -7.051 $ 
 & 
 $ -6.930 $ 
 & 
 $ -6.802 $ 
\\
4
 & 
[1/2]
 & 
 $ -8.346 $ 
 & 
 $ -7.690 $ 
 & 
 $ -7.440 $ 
 & 
 $ -7.236 $ 
 & 
 $ -7.075 $ 
 & 
 $ -6.956 $ 
 & 
 $ -6.834 $ 
\\
4
 & 
[0/3]
 & 
 $ -8.241 $ 
 & 
 $ -7.493 $ 
 & 
 $ -7.209 $ 
 & 
 $ -6.984 $ 
 & 
 $ -6.813 $ 
 & 
 $ -6.691 $ 
 & 
 $ -6.580 $ 
\\
\hline
\end{tabular}

\parbox{\captionwidth}{
\caption[]{\label{tab:NAxin}
  $\xi$ dependence of $A_{NA}^{(2)}|_{M_W=0}$.
  }}
\end{center}
\end{table}
}

\subsection{Subleading terms in $M_W$}

Let us now turn to the power-suppressed terms of order $(M_W^2/M_t^2)^n$
($n>0$). Both for the ${\cal O}(M_W^2/M_t^2)$ and ${\cal
  O}(M_W^4/M_t^4)$ contribution seven terms in the expansion in $z$
could be evaluated.

\subsubsection{Quadratic terms in $M_W$}
%
{\footnotesize
\begin{table}[ht]
\begin{center}
\begin{tabular}{|l|l||r|r||r|r||r|r||r|r|}
\hline
\multicolumn{2}{|c||}{ }
 & 
\multicolumn{2}{|c||}{ $ A_{l}^{(2)} $ }
 & 
\multicolumn{2}{|c||}{ $ A_{F}^{(2)} $ }
 & 
\multicolumn{2}{|c||}{ $ A_{A}^{(2)} $ }
 & 
\multicolumn{2}{|c|}{ $ A_{NA}^{(2)} $ }
\\\hline
input
 & 
P.A.
 & 
\multicolumn{1}{|c|}{ $ z $ }
 & 
\multicolumn{1}{|c||}{ $ \omega $ }
 & 
\multicolumn{1}{|c|}{ $ z $ }
 & 
\multicolumn{1}{|c||}{ $ \omega $ }
 & 
\multicolumn{1}{|c|}{ $ z $ }
 & 
\multicolumn{1}{|c||}{ $ \omega $ }
 & 
\multicolumn{1}{|c|}{ $ z $ }
 & 
\multicolumn{1}{|c|}{ $ \omega $ }
\\\hline
6
 & 
[3/2]
 & 
 --- 
 & 
 $ -1.016 $ 
 & 
 $ 0.09766 $ 
 & 
 $ 0.09769 $ 
 & 
 $ ^{(\star)} -2.721 $ 
 & 
 --- 
 & 
 $ 3.358 $ 
 & 
 --- 
\\
6
 & 
[2/3]
 & 
 $ ^{(\star)} -1.013 $ 
 & 
 $ -0.9459 $ 
 & 
 $ 0.09766 $ 
 & 
 $ 0.09769 $ 
 & 
 $ -2.747 $ 
 & 
 --- 
 & 
 $ 3.358 $ 
 & 
 --- 
\\\hline
7
 & 
[4/2]
 & 
 $ -0.9737 $ 
 & 
 $ -0.9773 $ 
 & 
 $ 0.09766 $ 
 & 
 $ 0.09764 $ 
 & 
 $ ^{(\star)} -2.677 $ 
 & 
 $ -2.794 $ 
 & 
 $ 3.353 $ 
 & 
 --- 
\\
7
 & 
[3/3]
 & 
 $ -0.9758 $ 
 & 
 $ -0.9676 $ 
 & 
 $ 0.09766 $ 
 & 
 --- 
 & 
 $ ^{(\star)} -2.677 $ 
 & 
 --- 
 & 
 $ 3.354 $ 
 & 
 --- 
\\
7
 & 
[2/4]
 & 
 $ -0.9652 $ 
 & 
 $ -1.122 $ 
 & 
 $ 0.09766 $ 
 & 
 $ 0.09764 $ 
 & 
 $ -2.770 $ 
 & 
 --- 
 & 
 $ 3.356 $ 
 & 
 --- 
\\
\hline
\end{tabular}

\parbox{\captionwidth}{
\caption[]{\label{tab:lallmb1}
  Pad\'e results for the ${\cal O}(\alpha_s^2)$ coefficients at ${\cal
    O}(M_W^2/M_t^2)$.  }}
\end{center}
\end{table}
}

The resulting Pad\'e values for the quadratic $M_W$
terms are listed in Table~\ref{tab:lallmb1}.
The light-fermion contribution is very stable and the maximal deviation from
the central value amounts to roughly 10\%.
Like in the case $M_W=0$, the Pad\'e approximants for $A_{F}^{(2)}$
--- both in $\omega$ and $z$ --- 
exhibit an impressive convergence.

For the abelian and non-abelian contribution
most of the \wpade s develop poles inside $|\omega|\leq 1$.  In
addition, while the \zpade s for $A_{NA}^{(2)}$ are very smooth, for
$A_A^{(2)}$ there are only two of them without poles within $|z|\le 1$.
However, all the poles of the other \zpade s approximately cancel
against zeros in the numerator. All relevant numbers are highly
consistent.

Therefore, the numbers in Table~\ref{tab:lallmb1} lead us to the following 
results:
\begin{eqnarray}
A_{A}^{(2)}\bigg|_{M_W^2} &=&  -2.73(6)
\,,
\nonumber\\
A_{NA}^{(2)}\bigg|_{M_W^2} &=&  3.356(3)
\,,
\nonumber\\
A_{l}^{(2)}\bigg|_{M_W^2} &=&  -1.0(1)
\,,
\nonumber\\
A_{F}^{(2)}\bigg|_{M_W^2} &=&  0.09766(3)
\,.
\label{eq:as2mw2}
\end{eqnarray}

\subsubsection{Quartic terms in $M_W$}
%
In Table~\ref{tab:lallmb2} the Pad\'e approximations for the quartic
terms in $M_W$ are listed.  The conformal mapping seems to spoil the
convergence property here, as all \wpade s develop poles within the unit
circle.  Moreover, in contrast to the constant and quadratic
corrections, for the $M_W^4$ terms it turns out that variations of
Eqs.~(\ref{eq:gamtwb}) and~(\ref{eq:Z2os}) (see the discussion in
Section~\ref{sec:method}) lead to results which lie outside the error
interval obtained from the numbers of Table~\ref{tab:lallmb2}.  So the
final numbers for the $M_W^4$ terms should only be considered as
estimates on their order of magnitude. They will be presented below,
with an artificially increased error of about 50\%.

{\footnotesize
\begin{table}[ht]
\begin{center}
\begin{tabular}{|l|l||r|r||r|r||r|r||r|r|}
\hline
\multicolumn{2}{|c||}{ }
 & 
\multicolumn{2}{|c||}{ $ A_{l}^{(2)} $ }
 & 
\multicolumn{2}{|c||}{ $ A_{F}^{(2)} $ }
 & 
\multicolumn{2}{|c||}{ $ A_{A}^{(2)} $ }
 & 
\multicolumn{2}{|c|}{ $ A_{NA}^{(2)} $ }
\\\hline
input
 & 
P.A.
 & 
\multicolumn{1}{|c|}{ $ z $ }
 & 
\multicolumn{1}{|c||}{ $ \omega $ }
 & 
\multicolumn{1}{|c|}{ $ z $ }
 & 
\multicolumn{1}{|c||}{ $ \omega $ }
 & 
\multicolumn{1}{|c|}{ $ z $ }
 & 
\multicolumn{1}{|c||}{ $ \omega $ }
 & 
\multicolumn{1}{|c|}{ $ z $ }
 & 
\multicolumn{1}{|c|}{ $ \omega $ }
\\\hline
6
 & 
[3/2]
 & 
 $ -1.290 $ 
 & 
 --- 
 & 
 $ 0.1474 $ 
 & 
 --- 
 & 
 --- 
 & 
 --- 
 & 
 $ 2.364 $ 
 & 
 --- 
\\
6
 & 
[2/3]
 & 
 $ -1.291 $ 
 & 
 --- 
 & 
 $ ^{(\star)} 0.1384 $
 & 
 --- 
 & 
 $ ^{(\star)} 4.606 $ 
 & 
 --- 
 & 
 $ 2.448 $ 
 & 
 --- 
\\\hline
7
 & 
[4/2]
 & 
 $ ^{(\star)} -1.268 $ 
 & 
 --- 
 & 
 $ 0.1479 $ 
 & 
 --- 
 & 
 $ ^{(\star)} 4.460 $ 
 & 
 --- 
 & 
 $ 2.489 $ 
 & 
 --- 
\\
7
 & 
[3/3]
 & 
 $ ^{(\star)} -1.278 $ 
 & 
 --- 
 & 
 $ 0.1477 $ 
 & 
 --- 
 & 
 $ ^{(\star)} 4.461 $ 
 & 
 --- 
 & 
 $ 2.459 $ 
 & 
 --- 
\\
7
 & 
[2/4]
 & 
 $ -1.305 $ 
 & 
 --- 
 & 
 $ 0.1489 $ 
 & 
 --- 
 & 
 --- 
 & 
 --- 
 & 
 $ 2.460 $ 
 & 
 --- 
\\
\hline
\end{tabular}

\parbox{\captionwidth}{
\caption[]{\label{tab:lallmb2}
  Pad\'e results for the ${\cal O}(\alpha_s^2)$ coefficients at ${\cal
    O}(M_W^4/M_t^4)$.  }}
\end{center}
\end{table}
}

While the leading two terms in $M_W$ do not contain logarithms of $M_W$,
the coefficients of $M_W^4$ develop linear logarithms of $M_W^2/M_t^2$.
For $A^{(1)}$ the coefficient of this logarithm is exactly reproduced by
the first two terms in the Taylor expansion around $z=0$ after setting
$z=1$; the higher order terms in $z$ vanish (a similar behaviour for the
logarithmic terms has already been observed in \cite{HarSeiSte97}).  The
phenomenon of a truncated series in $z$ for this coefficient also
appears for $A^{(2)}_{l,exp}$ and $A^{(2)}_{F,exp}$. According to the
discussion above, this strongly suggests that the logarithms of $M_W$ are
exactly recovered after setting $z=1$.  While for $A_l^{(2)}$ one 
arrives at $7/4\cdot \ln M_t^2/M_W^2$ for the term under
consideration, it sums up to zero for $A_F^{(2)}$.

For the abelian and non-abelian parts, the series in $z$ for the
coefficients of the $M_W^4/M_t^4\cdot \ln M_t^2/M_W^2$ term does not
seem to be truncated, but the Pad\'e analysis turns out to be very
stable. It can be found in Table~\ref{tab:ANAlmb}.

{\footnotesize
\begin{table}[ht]
\begin{center}
\begin{tabular}{|l|l||r|r||r|r|}
\hline
\multicolumn{2}{|c||}{ }
 & 
\multicolumn{2}{|c||}{ $ A_{A}^{(2)} $ }
 & 
\multicolumn{2}{|c|}{ $ A_{NA}^{(2)} $ }
\\\hline
input
 & 
P.A.
 & 
\multicolumn{1}{|c|}{ $ z $ }
 & 
\multicolumn{1}{|c||}{ $ \omega $ }
 & 
\multicolumn{1}{|c|}{ $ z $ }
 & 
\multicolumn{1}{|c|}{ $ \omega $ }
\\\hline
6
 & 
[3/2]
 & 
 $ 0.7811 $ 
 & 
 --- 
 & 
 $ -3.614 $ 
 & 
 --- 
\\
6
 & 
[2/3]
 & 
 $ 0.7829 $ 
 & 
 --- 
 & 
 $ -3.614 $ 
 & 
 --- 
\\\hline
7
 & 
[4/2]
 & 
 $ 0.7334 $ 
 & 
 --- 
 & 
 $ -3.626 $ 
 & 
 --- 
\\
7
 & 
[3/3]
 & 
 $ 0.7747 $ 
 & 
 --- 
 & 
 $ -3.618 $ 
 & 
 --- 
\\
7
 & 
[2/4]
 & 
 $ 0.7594 $ 
 & 
 --- 
 & 
 $ -3.622 $ 
 & 
 --- 
\\
\hline
\end{tabular}

\parbox{\captionwidth}{
\caption[]{\label{tab:ANAlmb}
  Pad\'e results for the coefficient of $M_W^4/M_t^4 \cdot \ln(M_W^2/M_t^2)$
  for $A^{(2)}_{A}$ and $A^{(2)}_{NA}$.  }}
\end{center}
\end{table}
}

Finally, the results for the quartic contributions in $M_W^4$ read:
\begin{eqnarray}
A_{A}^{(2)}\bigg|_{M_W^4} &=&  4.5(2.2) + 0.7(1)\,\ln\frac{M_t^2}{M_W^2}
\,,
\nonumber\\
A_{NA}^{(2)}\bigg|_{M_W^4} &=&  2.4(1.2) - 3.62(1)\,\ln\frac{M_t^2}{M_W^2}
\,,
\nonumber\\
A_{l}^{(2)}\bigg|_{M_W^4} &=&  -1.3(7) + \frac{7}{4}\ln\frac{M_t^2}{M_W^2}
\,,
\nonumber\\
A_{F}^{(2)}\bigg|_{M_W^4} &=&  0.15(5)
\,.
\label{eq:as2mw4}
\end{eqnarray}

\subsection{Results at ${\cal O}(\alpha_s^2)$}

In this subsection we finally present the numerical corrections for
$\Gamma(t\to Wb)$ at ${\cal O}(\alpha_s^2)$. Simply using the results
derived above according to (\ref{eq:coldec}) and linearly adding the
errors one would certainly overestimate the total uncertainty. 
It is more promising to add up the expansions in $z$ for different colour
structures and to perform the Pad\'e procedure afterwards. 
The corresponding Pad\'e approximations for the case $M_W=0$ are shown in 
Table~\ref{tab:topmb0}. Table~\ref{tab:topmb12lmb} contains the 
$M_W$-suppressed terms. The behaviour is similar to the approach where the
individual colour structures are treated separately. In the case of vanishing
$W$ boson mass the Pad\'e results both with and without conformal mapping are
highly stable.  For the $M_W^2$- and $M_W^4$- terms, however, many of the 
\wpade s develop poles inside the unit circle. The result we deduce from 
Tables~\ref{tab:topmb0} and \ref{tab:topmb12lmb} reads:
\begin{eqnarray}
A^{(2)} &=& -16.7(8)
       + 5.4(4) \, \frac{M_W^2}{M_t^2}
       + \frac{M_W^4}{M_t^4}\left(
         11.4(5.0) - 7.3(1) \, \ln{M_t^2\over M_W^2}
        \right) 
\label{eq:A2comb}
\,.
\end{eqnarray}

The leading-order result is in very good agreement with the one
of~\cite{CzaMel98} which reads $-16.7(5)$. Note again that the error
estimate for the $M_W^4$ term is rather conservative.
{\footnotesize
\begin{table}[ht]
\begin{center}
\begin{tabular}{|l|l||r|r||r|r|}
\hline
\multicolumn{2}{|c||}{ }
 & 
\multicolumn{2}{|c||}{ $ A^{(2)} $ }
\\\hline
input
 & 
P.A.
 & 
\multicolumn{1}{|c|}{ $ z $ }
 & 
\multicolumn{1}{|c||}{ $ \omega $ }
\\\hline
6
 & 
[3/2]
 & 
 $ -16.83 $ 
 & 
 $ -16.73 $ 
\\
6
 & 
[2/3]
 & 
 $ -16.84 $ 
 & 
 --- 
\\\hline
7
 & 
[4/2]
 & 
 $ -16.89 $ 
 & 
 $ -16.49 $ 
\\
7
 & 
[3/3]
 & 
 $ -16.91 $ 
 & 
 $ -15.85 $ 
\\
7
 & 
[2/4]
 & 
 $ -16.90 $ 
 & 
 $ -16.79 $ 
\\\hline
8
 & 
[4/3]
 & 
 $ -16.95 $ 
 & 
 --- 
\\
8
 & 
[3/4]
 & 
 $ -16.95 $ 
 & 
 $ -16.83 $ 
\\
\hline
\end{tabular}

\parbox{\captionwidth}{
\caption[]{\label{tab:topmb0}
  Pad\'e results for $A^{(2)}|_{M_W=0}$.
  }}
\end{center}
\end{table}
}
{\footnotesize
\begin{table}[ht]
\begin{center}
\begin{tabular}{|l|l||r|r||r|r||r|r|r|}
\hline
\multicolumn{2}{|c||}{ }
 & 
\multicolumn{2}{|c||}{ $ M_W^2 $ }
 & 
\multicolumn{2}{|c||}{ $ M_W^4 $ }
 & 
\multicolumn{2}{|c|}{ $ M_W^4 \ln (M_t^2/M_W^2) $ }
\\\hline
input
 & 
P.A.
 & 
\multicolumn{1}{|c|}{ $ z $ }
 & 
\multicolumn{1}{|c||}{ $ \omega $ }
 & 
\multicolumn{1}{|c|}{ $ z $ }
 & 
\multicolumn{1}{|c||}{ $ \omega $ }
 & 
\multicolumn{1}{|c|}{ $ z $ }
 & 
\multicolumn{1}{|c|}{ $ \omega $ }
\\\hline
6
 & 
[3/2]
 & 
 $ ^{(\star)} 5.774 $
 & 
 --- 
 & 
 --- 
 & 
 --- 
 & 
 $ -7.275 $ 
 & 
 --- 
\\
6
 & 
[2/3]
 & 
 $ 5.340 $ 
 & 
 --- 
 & 
 --- 
 & 
 --- 
 & 
 $ -7.281 $ 
 & 
 --- 
\\\hline
7
 & 
[4/2]
 & 
 $ 5.197 $ 
 & 
 --- 
 & 
 $ 11.54 $ 
 & 
 --- 
 & 
 $ -7.354 $ 
 & 
 --- 
\\
7
 & 
[3/3]
 & 
 $ 5.198 $ 
 & 
 $ 5.338 $ 
 & 
 --- 
 & 
 --- 
 & 
 $ -7.304 $ 
 & 
 --- 
\\
7
 & 
[2/4]
 & 
 $ 5.280 $ 
 & 
 --- 
 & 
 $ 11.21 $ 
 & 
 --- 
 & 
 $ -7.319 $ 
 & 
 --- 
\\
\hline
\end{tabular}

\parbox{\captionwidth}{
\caption[]{\label{tab:topmb12lmb}
  Pad\'e results for the coefficients of $M_W^2/M_t^2$, $M_W^4/M_t^4$, and
  $M_W^4/M_t^4 \cdot \ln{M_t^2/M_W^2}$ of $A^{(2)}$.
  }}
\end{center}
\end{table}
}

Finally we are in the position to write down the decay rate
of the top quark up to order ${\cal O}(\alpha_s^2 M_W^4/M_t^4)$.
Using the exact results at Born level and at order $\alpha_s$ in combination
with Eq.~(\ref{eq:A2comb}) leads to:
\begin{eqnarray}
\Gamma(t\to bW) &=&
\Gamma_0\left[
  0.8852
  - 2.220 \, \frac{\alpha_s}{\pi}
  - 15.6(1.1) \left(\frac{\alpha_s}{\pi}\right)^2
  +\ldots
\right]
\nonumber\\
&=&
0.788(1)\,\Gamma_0
\label{eq:twbnumos}
\,,
\end{eqnarray}
where the values $M_t=175$~GeV, $M_W=80.4$~GeV and $\alpha_s(M_t^2)=0.11$
have been assumed.
For $M_W=0$ the second order QCD corrections amount to roughly 2\%.
Both at order $\alpha_s$ (see Eq.~(\ref{eq:a1})) and $\alpha_s^2$ the 
$M_W$ mass corrections ``screen'' the leading order terms, i.e.,
they arise with negative sign.
Whereas the quadratic and quartic corrections at ${\cal O}(\alpha_s)$
turn out to be 16\% and 3\% w.r.t. to the massless result,
they amount to roughly
7\% and 1\% at ${\cal O}(\alpha_s^2)$, respectively.

Using the ${\cal O}(\alpha_s^2)$ relation between $M_t$ and 
the $\overline{\rm MS}$ mass $m_t(\mu)$~\cite{GraBroGraSch90} one 
may express the result in terms of $m_t \equiv m_t(m_t)$:
\begin{eqnarray}
\lefteqn{\bar{\Gamma}(t\to bW) = }\nonumber\\&=&
\bar\Gamma_0\bigg\{
  1 - 3{M_W^4\over m_t^4} + 2{M_W^6\over m_t^6}
  + \frac{\alpha_s}{\pi}\left[1.28 + 2\,{M_W^2\over m_t^2} +
  {M_W^4\over m_t^4}
  \left(9.16 - 2\,\ln{m_t^2\over M_W^2}\right) + \ldots\right]
    \nonumber\\&&
  + \left(\frac{\alpha_s}{\pi}\right)^2\,\left[2.5(8) + 8.1(4)\,{M_W^2\over
    m_t^2} + {M_W^4\over m_t^4}\,\left(18.6(5.0) - 4.6(1)\,\ln{m_t^2\over
    M_W^2}\right) + \ldots\right]
\bigg\}
\nonumber\\
&=& \bar\Gamma_0\left[0.8576 + 1.98\,{\alpha_s\over \pi} +
5.0(1.2)\,\left({\alpha_s\over \pi}\right)^2 + \ldots \right]\nonumber\\
&=& 0.933(1)\,\bar\Gamma_0
\label{eq:twbnumms}
\,,
\end{eqnarray}
where $\bar\Gamma_0 = G_Fm_t^3|V_{tb}|^2/(8\pi\sqrt{2})$,
$\alpha_s(m_t^2) = 0.11$, and $m_t=165$~GeV has been chosen. At ${\cal
  O}(\alpha_s)$ the exact result is used after the second equality sign.
As expected, the convergence of the perturbative series is better in
this case than in Eq.~(\ref{eq:twbnumos}).

Let us finally compare the full ${\cal O}(\alpha_s^2)$ result with the 
BLM~\cite{BroLepMac83} contributions. They are obtained by replacing the 
number of light fermions in Eq.~(\ref{eq:2loopdef}) by 
$-({33 \over 2} - n_l)$ and neglecting contributions from other colour 
factors. The result reads:
\begin{eqnarray}
A^{(2),\mbox{\footnotesize BLM}} &=& -21.92 + 7.7(8) \, \frac{M_W^2}{M_t^2}
       + \frac{M_W^4}{M_t^4}\left(
         9.9(2) - {161 \over 12} \, \ln{M_t^2\over M_W^2}
        \right)
\label{eq:A2BLM}
\,.
\end{eqnarray}
For vanishing $W$ boson mass the difference to the complete order
$\alpha_s^2$ result amounts to roughly 24\% and the ${\cal O}(M_W^2)$
term is off by almost 50\%.  The order of magnitude for the quartic term
in $M_W$ is reproduced correctly, but the logarithmic term differs from
(\ref{eq:A2comb}) by almost a factor of 2.


\subsection{Estimate for $b\to ul\bar\nu$ and $\mu \to e\nu_\mu\bar\nu_e$
  \label{sec:bulnu}}
%
As it was pointed out in \cite{CzaMel98,Melpriv}, the results for top
decay may be used to estimate also the QCD corrections to the
semi-leptonic decay of the bottom quark.  The ${\cal O}(\alpha_s^2)$
corrections to this process have been obtained recently \cite{Rit99} by
computing four-loop on-shell diagrams. Nevertheless, using the results
of the previous sections we may also derive an approximation to this
quantity, in this way verifying the consistency of the results of
\cite{Rit99}, \cite{CzaMel98} and the present paper.

The decay rate for $b\to ul\bar\nu$ can be expressed as
\begin{equation}\label{eq:mudef}
\Gamma(b\to ul\bar\nu) = \Gamma_b^{(0)} + {\alpha_s\over
  \pi}C_F\Gamma_b^{(1)} + \left({\alpha_s\over
  \pi}\right)^2\Gamma_b^{(2)} + \ldots
\end{equation}
with
\begin{equation}\label{eq:gambdy}
\Gamma_b^{(i)} = 2\,\Gamma_b^{(0)}\,\int_0^1\dd y A^{(i)}(y)\,,\qquad
\Gamma_b^{(0)} = {G_F^2|V_{ub}|^2M_b^5\over 192\pi^3}\,,
\end{equation}
where $M_b$ is the on-shell bottom quark mass and $V_{ub}$ is the CKM
matrix element for $b\to u$ transitions.
The relation to the top decay rate is established through
$A^{(i)}(M_W^2/M_t^2) \equiv A^{(i)}$, with the $A^{(i)}$ defined in
Eq.~(\ref{eq:coldec}). Assuming that the functions
\begin{equation}
\hat A^{(i)}(y) \equiv {A^{(i)}(y)\over A^{(0)}(y)}
\end{equation}
are smooth within $0<y<1$, one may approximate them by their first few terms
in the expansion around $y=0$:
\begin{equation}
\hat A^{(i)}(y) = \sum_{n\ge 0}\left(a_n^{(i)} + a_{L,n}^{(i)}\,\log
  y\right) y^n\,,
\end{equation}
which leads to
\begin{equation}\label{eq:gamg0}
\Gamma_b^{(i)} = 2\Gamma_b^{(0)}\sum_{n\ge 0}\left(
  a_n^{(i)}\,\int_0^1\dd y A^{(0)}(y) y^n + a_{L,n}^{(i)}\,
  \int_0^1\dd y A^{(0)}(y) y^n \log y\right)\,.
\end{equation}
For example, in $0^{th}$ approximation, one finds ($a_{L,0}^{(i)} = 0$
  for $i=0,1,2$)
\begin{equation}
\Gamma_b^{(i)} = a_0^{(i)}\Gamma_b^{(0)} = A^{(i)}(0)\,
\Gamma_b^{(0)}\,.
\end{equation}
However, $\hat A^{(i)}(y)$ is not really smooth in general. In fact,
$\hat A^{(1)}(y)$ has a singularity at $y=1$ which spoils convergence of
the expansion in $y$. On the other hand, $A^{(i)}(y) = A^{(0)}(y)\,\hat
A^{(i)}(y)$ itself is finite for $y=1$. Thus, if a larger number of
terms in $y$ is included, it is more promising to use directly
Eq.~(\ref{eq:gambdy}) and expand the full integrand around small $y$.
This is demonstrated in the case of $\Gamma_b^{(1)}$ in
Table~\ref{tab:bdecas1} where both approaches are compared including
successively higher powers in $y$.

{\footnotesize
\begin{table}
  \begin{center}
    \begin{tabular}{|l||c|c|}
\hline
 & 
\multicolumn{2}{|c|}{$ \Gamma_b^{(1)}/\Gamma_b^{(0)} $}
\\\hline
 $ n $ 
 & 
 $(a)$
 & 
 $(b)$
\\\hline
0
 & 
$ -2.040 $
 & 
$ -4.080 $
\\
1
 & 
$ -1.590 $
 & 
$ -2.580 $
\\
2
 & 
$ -2.030 $
 & 
$ -0.3333 $
\\
3
 & 
$ -1.495 $
 & 
$ -0.9843 $
\\
4
 & 
$ -2.093 $
 & 
$ -1.744 $
\\
5
 & 
$ -1.327 $
 & 
$ -1.810 $
\\
6
 & 
$ -2.371 $
 & 
$ -1.818 $
\\
7
 & 
$ -0.8998 $
 & 
$ -1.818 $
\\
8
 & 
$ -3.071 $
 & 
$ -1.816 $
\\
9
 & 
$ 0.2045 $
 & 
$ -1.815 $
\\
10
 & 
$ -4.881 $
 & 
$ -1.814 $
\\\hline
 exact: 
 & 
\multicolumn{2}{|c|}{$ - 1.810 $}
\\
\hline
    \end{tabular}
\parbox{\captionwidth}{
\caption[]{\label{tab:bdecas1}
  Estimates for $\Gamma_b^{(1)}$ using $(a)$ Eq.~(\ref{eq:gamg0}),
  and $(b)$ Eq.~(\ref{eq:gambdy}) with the full integrand replaced
  by its expansion around small $y$. $n$ is the order of the expansion
  in $y$ that was used as input.
}}
  \end{center}
\end{table}
}

One can see that the approach using Eq.~(\ref{eq:gamg0}) provides
reasonable estimates for $n\lsim 4$, where, on the other hand, the
results obtained by a naive expansion of the integrand in
Eq.~(\ref{eq:gambdy}) are unsatisfactory. For $n>4$, however, the
situation becomes opposite: The more terms in $y$ are included, the
better is the approximation using the latter method. The method using
Eq.~(\ref{eq:gamg0}) becomes very unstable.

The same procedure will now be applied at ${\cal O}(\alpha_s^2)$.  As
input we use the numbers of Eq.~(\ref{eq:A2comb}) and subtract the
values for $A_l^{(2)}$ as given in Eqs.~(\ref{eq:resal2}),
(\ref{eq:as2mw2}), and (\ref{eq:as2mw4}), multiplied by $C_F T$,
according to the transition from $n_l=5$ for top decay to $n_l=4$ for
bottom decay.  Only the central values from these equations will be
used, the errors will be suppressed.  The results are shown in
Table~\ref{tab:bdecas2}.  One can see that both approaches lead to
results that are fairly consistent with the exact number obtained in
\cite{Rit99}.

Along the same line of reasoning we may derive an estimate for the
${\cal O}(\alpha^2)$-corrections to the decay rate of the muon. Applying
the obvious modifications to the notation of Eqs.~(\ref{eq:mudef}) and
(\ref{eq:gambdy}) and using the results of Eqs.~(\ref{eq:resal2}),
(\ref{eq:as2mw2}), and (\ref{eq:as2mw4}), we find the numbers given in
Table~\ref{tab:mudec}. Only the method according to Eq.~(\ref{eq:gamg0})
has been applied. Again these results agree nicely with the exact
results obtained in \cite{RitStu99}.

This agreement can be considered as a non-trivial check 
of the results of \cite{Rit99,RitStu99} and the ones obtained in this paper.
{\footnotesize
\begin{table}
  \begin{center}
    \begin{tabular}{|l||c|c|}
\hline
 & 
\multicolumn{2}{|c|}{$ \Gamma_b^{(2)}/\Gamma_b^{(0)} $}
\\\hline
 $ n $ 
 & 
 $ (a) $
 & 
 $ (b) $
\\\hline
0
 & 
 $ -18.6 $ 
 & 
$ -37.1 $
\\
1
 & 
 $ -16.8 $ 
 & 
$ -31.2 $
\\
2
 & 
 $ -23.4 $ 
 & 
$ -24.9 $
\\\hline
 exact~\cite{Rit99}: 
 & 
\multicolumn{2}{|c|}{$ - 21.3 $}
\\
\hline
\end{tabular}
\parbox{\captionwidth}{
\caption[]{\label{tab:bdecas2}
  Estimates for $\Gamma_b^{(2)}$. Same notation as in
  Table~\ref{tab:bdecas1}.}}
\end{center}
\end{table}
}

{\footnotesize
  \begin{table}
  \begin{center}
    \begin{tabular}{|l||c|c|c||c|}
\hline
 & 
\multicolumn{4}{|c|}{ $ \Gamma_\mu^{(2)}/\Gamma_\mu^{(0)} $ }
\\\hline
 $ n $ 
 & 
 $ \gamma\gamma $ 
 & 
 elec 
 & 
 muon 
 & 
 $ \Sigma $ 
\\\hline
0
 & 
 $ 3.20 $ 
 & 
 $ 2.80 $ 
 & 
 $ -0.0636 $ 
 & 
 $ 5.94 $ 
\\
1
 & 
 $ 2.38 $ 
 & 
 $ 2.50 $ 
 & 
 $ -0.0343 $ 
 & 
 $ 4.85 $ 
\\
2
 & 
 $ 4.33 $ 
 & 
 $ 3.61 $ 
 & 
 $ -0.0397 $ 
 & 
 $ 7.90 $ 
\\\hline
 exact~\cite{RitStu99}: 
 & 
 $ 3.56 $ 
 & 
 $ 3.22 $ 
 & 
 $ -0.0364 $ 
 & 
 $ 6.74 $ 
\\
\hline
    \end{tabular}
\parbox{\captionwidth}{
\caption[]{\label{tab:mudec}
  Estimates for $\Gamma_\mu^{(2)}$. ``$\gamma\gamma$'' denotes the
  purely photonic corrections, ``elec'' and ``muon'' the ones involving
  electron and muon loops, respectively. (``elec'' also includes real
  emission of an electron-positron pair, of course.)}}
\end{center}
\end{table}
}


\section{Conclusions}

In this work QCD corrections of order $\alpha_s^2$ to the decay of the
top quark into a $W$ boson and a bottom quark have been considered.
Since the exact treatment of the contributing Feynman diagrams is currently
out of question, the calculation has been reduced to the evaluation of
moments.  The physical limit is obtained via conformal mapping and
Pad\'e approximation. The existing results in the limit of a massless
$W$ boson could be confirmed and new terms of order $M_W^2/M_t^2$ and
$M_W^4/M_t^4$ were obtained.  Numerically it turns out that these
power-suppressed terms are rather small.  Assuming similar convergence
properties concerning $M_W$ for the one- and two-loop corrections we can
conclude that the ${\cal O}(\alpha_s^2)$ corrections for $\Gamma(t\to
Wb)$ are well under control, including finite $W$-mass effects.

The approach used in this article for the evaluation of the diagrams can
certainly be carried over to other interesting physical problems, e.g.,
semileptonic bottom quark decays or muon decay as indicated in
Section~\ref{sec:bulnu}.  In this paper the reliability of the method
has been demonstrated by a comparison with a completely different
approach.


\section*{Acknowledgments} 

We would like to thank A.~Czarnecki and K.~Melnikov for encouragement,
numerous fruitful discussions, and providing us with a copy of
\cite{CzaMel98_2} before its publication.
We are indebted to J.H.~K\"uhn for valuable comments, encouraging
discussions, and careful reading of the manuscript.
This work was supported by DFG under Contract Ku 502/8-1, the {\it
  Graduiertenkolleg ``Elementarteilchenphysik an Beschleunigern''}, the
{\it DFG-Forschergruppe ``Quantenfeldtheorie, Computeralgebra und
  Monte-Carlo-Simulationen''} and
the {\it Landesgraduiertenf\"orderung} at the University of Karlsruhe, 
and the {\it Schweizer Nationalfond}.


\def\app#1#2#3{{\it Act.~Phys.~Pol.~}{\bf B #1} (#2) #3}
\def\apa#1#2#3{{\it Act.~Phys.~Austr.~}{\bf#1} (#2) #3}
\def\cmp#1#2#3{{\it Comm.~Math.~Phys.~}{\bf #1} (#2) #3}
\def\cpc#1#2#3{{\it Comp.~Phys.~Commun.~}{\bf #1} (#2) #3}
\def\epjc#1#2#3{{\it Eur.\ Phys.\ J.\ }{\bf C #1} (#2) #3}
\def\fortp#1#2#3{{\it Fortschr.~Phys.~}{\bf#1} (#2) #3}
\def\ijmpc#1#2#3{{\it Int.~J.~Mod.~Phys.~}{\bf C #1} (#2) #3}
\def\ijmpa#1#2#3{{\it Int.~J.~Mod.~Phys.~}{\bf A #1} (#2) #3}
\def\jcp#1#2#3{{\it J.~Comp.~Phys.~}{\bf #1} (#2) #3}
\def\jetp#1#2#3{{\it JETP~Lett.~}{\bf #1} (#2) #3}
\def\mpl#1#2#3{{\it Mod.~Phys.~Lett.~}{\bf A #1} (#2) #3}
\def\nima#1#2#3{{\it Nucl.~Inst.~Meth.~}{\bf A #1} (#2) #3}
\def\npb#1#2#3{{\it Nucl.~Phys.~}{\bf B #1} (#2) #3}
\def\nca#1#2#3{{\it Nuovo~Cim.~}{\bf #1A} (#2) #3}
\def\plb#1#2#3{{\it Phys.~Lett.~}{\bf B #1} (#2) #3}
\def\prc#1#2#3{{\it Phys.~Reports }{\bf #1} (#2) #3}
\def\prd#1#2#3{{\it Phys.~Rev.~}{\bf D #1} (#2) #3}
\def\pR#1#2#3{{\it Phys.~Rev.~}{\bf #1} (#2) #3}
\def\prl#1#2#3{{\it Phys.~Rev.~Lett.~}{\bf #1} (#2) #3}
\def\pr#1#2#3{{\it Phys.~Reports }{\bf #1} (#2) #3}
\def\ptp#1#2#3{{\it Prog.~Theor.~Phys.~}{\bf #1} (#2) #3}
\def\sovnp#1#2#3{{\it Sov.~J.~Nucl.~Phys.~}{\bf #1} (#2) #3}
\def\tmf#1#2#3{{\it Teor.~Mat.~Fiz.~}{\bf #1} (#2) #3}
\def\yadfiz#1#2#3{{\it Yad.~Fiz.~}{\bf #1} (#2) #3}
\def\zpc#1#2#3{{\it Z.~Phys.~}{\bf C #1} (#2) #3}
\def\ppnp#1#2#3{{\it Prog.~Part.~Nucl.~Phys.~}{\bf #1} (#2) #3}
\def\ibid#1#2#3{{ibid.~}{\bf #1} (#2) #3}

\end{document}